\documentstyle[11pt]{article}

\newcommand{\BE}{\begin{equation}}
\newcommand{\EE}{\end{equation}}
\newcommand{\BA}{\begin{eqnarray}}
\newcommand{\EA}{\end{eqnarray}}
\newcommand{\half}{{\scriptstyle{\frac{1}{2}}}}
\begin{document}
\begin{titlepage}
\begin{flushright}
\hspace*{8.5cm} {\small DE-FG05-92ER40717-15 \\
hep-ph/9507408}
\end{flushright}

\vspace*{24mm}
\begin{center}
               {\Huge \bf The $\lambda\Phi^4$ Miracle: }\\
\vspace*{6mm}
        {\LARGE \bf Lattice data and the zero-point potential}

\vspace*{25mm}
{\Large P. M. Stevenson}

\vspace*{5mm}
{\large T.~W.~Bonner Laboratory, Physics Department \\
Rice University, Houston, TX 77251, USA}

\vspace{20mm}
{\bf Abstract:}
\end{center}

\par
    Recent lattice data for the effective potential of $\lambda \Phi^4$
theory fits the massless one-loop formula with amazing precision.
Any corrections are at least 100 times smaller than is reasonable,
perturbatively.  This is strong evidence for the ``exactness
conjecture'' of Consoli and Stevenson.

\end{titlepage}

\setcounter{page}{1}

     {\bf 1.}
     This note is to draw attention to a very striking result obtained
recently from lattice simulations of the four-dimensional $\lambda \Phi^4$
theory \cite{aac,cc,cccf}.  At certain parameter values the shape of the
effective potential agrees with the (unrenormalized, massless) one-loop
formula to amazing precision:  no deviations are found at the $10^{-4}$
level, {\it even though perturbatively one would expect corrections of
order a few percent}.  This result is predicted by the picture of Consoli
and Stevenson \cite{cs,csz}; from any conventional viewpoint the result
is an inexplicable miracle.

     My aim here is simply to bring out the main point briefly and
dramatically.  For a thorough discussion, see Refs. \cite{aac,cccf}.
It will suffice to study just one set of data points (obtained by
Cea and Cosmai \cite{cc} and published in Table 3 of Ref. \cite{aac}).
Other data sets amply comfirm the result.  My analysis is extremely
simple, and the sceptical reader may check it in five minutes with a
pocket calculator.

     The lattice simulations used the discretized Euclidean action:
\BE
   S_J = \sum_x \left[ \frac{1}{2} \sum_{\hat{\mu}}
(\Phi_{x+\hat{\mu}} - \Phi_x)^2 + \frac{1}{2} m_B^2 \Phi_x^2
+ \frac{1}{4!} \lambda_B \Phi_x^4 - J \Phi_x \right]
\EE
(where $x$ denotes a generic lattice site and $x+\hat{\mu}$ its nearest
neighbours).  The data were taken on a $16^4$ lattice at $\lambda_B = 6$
and $m_B^2 = -0.45$.  (All numerical values are in lattice units:
$\lambda_B = 6 \lambda_0$, and $m_B^2 = r_0$ in the notation of
\cite{aac,cccf}.)  The value of $m_B^2$ was chosen to yield the massless
case, in the sense of Coleman and Weinberg \cite{cw}.  The expectation
value, $\phi_B(J) \equiv \langle \Phi \rangle_J$ of the bare field was
measured for various input values of the constant source $J$.  See
the first two columns of Table 1 below.  Viewed in reverse, these data
give the value of $J$ at a particular $\phi_B$ value --- and $J(\phi_B)$
is ${\rm d} V_{\rm eff}/{\rm d} \phi_B$ \cite{cw}.

\begin{center}

\begin{tabular}{l l|l}
$\hspace*{1em} J$ & \hspace*{3.5em} $\phi_B$ & \hspace*{2em} $R(J, \phi_B)$ \\
\hline
$0.100$ \quad & $0.506086 \pm(0.91 \times 10^{-4})$ \hspace*{1em} &
 \quad $17.562 \pm 0.010$ \\
$0.125$       & $0.543089 \pm(0.82 \times 10^{-4})$       &
 \quad $17.543 \pm 0.008$ \\
$0.150$       & $0.575594 \pm(0.82 \times 10^{-4})$       &
 \quad $17.552 \pm 0.008$ \\
$0.200$       & $0.630715 \pm(0.62 \times 10^{-4})$       &
 \quad $17.549 \pm 0.005$ \\
$0.300$       & $0.717585 \pm(0.52 \times 10^{-4})$       &
 \quad $17.548 \pm 0.004$ \\
$0.400$       & $0.786503 \pm(0.44 \times 10^{-4})$       &
 \quad $17.552 \pm 0.003$ \\
$0.500$       & $0.844473 \pm(0.41 \times 10^{-4})$       &
 \quad $17.552 \pm 0.003$ \\
$0.600$       & $0.894993 \pm(0.39 \times 10^{-4})$       &
 \quad $17.551 \pm 0.002$ \\
$0.700$       & $0.940074 \pm(0.37 \times 10^{-4})$       &
 \quad $17.550 \pm 0.002$ \\

\end{tabular}

The lattice data [2,1] and the ratio
$R(J,\phi_B) \equiv \phi_B^3 \ln (\phi_B^2/v_B^2)/J$.

\end{center}

     [Note that the lattice calculation is unable to access the small $J$
region.  The statistical errors increase as $J$ decreases, and finite-volume
effects become large \cite{aac,cccf}.  This prevents one from directly
accessing the region $\phi_B \approx v_B$ as one would ideally like.]

     The ``miracle'' is the extraordinary precision with which these data
fit the simple formula
\BE
\label{form}
   J(\phi_B) = \alpha \phi_B^3 \ln(\phi_B^2/v_B^2)
\EE
(where $\alpha$ and $v_B$ are constants).  To demonstrate this fact
one could plot $J/\phi_B^3$ versus $\ln \phi_B^2$.  An even simpler
task, if one uses the value $v_B = 5.783 \times 10^{-4}$ obtained in Ref.
\cite{aac}'s fit, is to check that the ratio
$R(J,\phi_B) \equiv \phi_B^3 \ln (\phi_B^2/v_B^2)/J$ is constant.  This
is done in the third column of Table 1.  The constancy of $R$ is evident.
Fig. 1 shows this graphically, on a fine scale.  Each data point lies
within $1.1 \sigma$ of a common value, $R = 17.551$.  The
individual statistical errors are between $0.06\%$ and $0.01\%$.  There
is no sign of any systematic deviation from constancy at the $0.01\%$ level.

     {\bf 2.}
     This is the empirical result.  It accords beautifully with the
picture of Ref. \cite{cs,csz}, in which the form of the effective
potential is given exactly, in the continuum limit, by the ``zero-point
potential'' (ZPP) --- the classical potential plus the zero-point energy
of free-field fluctuations.  In the ``massless'' (or ``classically
scale-invariant'') case this gives
\BE
V_{\rm eff} \propto \phi_B^4 (\ln(\phi_B^2/v_B^2) - \half ).
\EE
The value of the constant of proportionality here is not physically
significant, since $\phi_B$ has to be renormalized.  Differentiating
this formula yields (\ref{form}).

      In the general case \cite{csz} $V_{\rm eff}$ also has a
$\phi_B^2$ term, giving an extra linear term in the expression for $J$.
This term is absent in the ``massless'' case, and  Ref. \cite{aac}
found that, at $\lambda_B=6$, one needs $m_B^2 = -0.45$ to pick out
this special case.  This finding is completely consistent with Brahm's
analysis \cite{brahm} of independent lattice data.  [Ref. \cite{cccf}
has taken data, at $\lambda_B = 3$, using Brahm's central value
$m_B^2 = -0.2279$ to avoid any possible bias when selecting the
``massless'' case.  The agreement with (\ref{form}) is equally
spectacular.]

     {\bf 3.}
     Now let us consider whether the results can be explained by
conventional theoretical ideas.  The loop expansion (after mass
renormalization by normal ordering, but before coupling-constant
renormalization) would give \cite{cw}
\BE
  V_{\rm eff} = \frac{\lambda_B}{4!} \phi_B^4 +
\frac{\lambda_B^2 \phi_B^4} {256 \pi^2}
\left( \ln \left( \frac{\half \lambda_B \phi_B^2}{\Lambda^2} \right)
- \frac{1}{2} \right) + \ldots,
\EE
where $\Lambda$ is the ultraviolet cutoff.   Differentiating and dividing
by $\phi_B^3$ yields, at $\lambda_B = 6$:
\BE
   J/\phi_B^3 = 1 + \frac{9}{16 \pi^2}
\ln \left( \frac{\half \lambda_B \phi_B^2}{\Lambda^2} \right)
+ \ldots .
\EE
If the ``$\mbox{} + \ldots$'' (2-loop and higher) terms were negligible,
then one could absorb the leading term, 1, into the scale of the
logarithm and obtain Eq. (\ref{form}) with
$\alpha^{-1} = 16 \pi^2/9 = 17.546$, close to the empirical value $17.551$.
[One could actually adjust $v_B$ to remove this small discrepancy.
Using $v_B = 5.7943 \times 10^{-4}$ (see \cite{aac}, Table 4)
leads to a figure virtually identical to Fig. 1 except for a shift in
the vertical scale.]

      The problem is that, from a conventional viewpoint, we do
{\it not} expect the ``$\mbox{} + \ldots$'' terms to be negligible.
For the quoted data, $J/\phi_B^3$ differs by roughly 20\% from its
leading-order value, 1.  This means that the one-loop correction
is about 20\%, and since the leading logs recur in a geometric series,
one would naturally expect the 2-loop term to be of order $(20\%)^2 = 4\%$.
Thus, the ratio $R(J, \phi_B)$ ought to show a systematic deviation from
constancy at the level of a few percent.  In fact there is no sign of
any such deviation at the $0.01\%$ level.

     The preceding argument ignored renormalization.  Conventional
renormalization introduces a renormalized coupling constant
$\lambda_R(\mu) = \lambda_B (1-b_0 \lambda_B \ln(\Lambda/\mu) + \ldots )$,
where $b_0 = 3/(16 \pi^2)$.  ``RG-improvement'' tells us (i) to
re-sum the series of leading logs (i.e., $(1-x+\ldots) \to 1/(1+x) + \ldots$,
where $x=b_0 \lambda_B \ln(\Lambda/\mu)$, and (ii) to choose the
renormalization scale $\mu$ to be of order $\phi$ (there is no distinction
between $\phi_B$ and $\phi_R$ at this level).  The ``RG-improved'' result
at the one-loop level is then basically the classical result with
$\lambda_B$ replaced by the running coupling constant \cite{cw}:
\BE
  \lambda_R(\phi) \equiv \frac{\lambda_R{\scriptstyle{(M)}}}
{1 - \frac{3 \lambda_R(M)}{32 \pi^2} \ln \frac{\phi^2}{M^2}},
\EE
where $M$ is some arbitary renormalization scale.
($\lambda_R({\scriptstyle{M \approx \Lambda}})$ is $\lambda_B$.)
This would predict
\BE
\label{ll}
J = \lambda_R(\phi) \phi^3 /6.
\EE
One can re-arrange this to yield
\BE
\ln \phi + \frac{8 \pi^2}{9} \frac{\phi^3}{J} = \mbox{\rm const.}
\EE
Computing the left-hand side from the data gives values that steadily
decrease from $10.691 \pm 0.006$ to $10.350 \pm 0.001$.  The deviation
from constancy is a few percent --- but that is much greater than the
errors allow.  [Ref. \cite{cccf} has tried fits to the full 2-loop formula,
including an adjustable mass parameter; the $\chi^2/{\rm d.o.f}$ is
still very poor, $152/14$.]  In one sense, there is nothing wrong because one
expects corrections of relative order $\lambda_R/(16 \pi^2)$; a few percent.
However, the point remains that the precise agreement with Eq. (\ref{form})
is completely baffling from a conventional viewpoint.  Why does the
``naive one-loop'' result fit the data perfectly at the $0.01 \%$ level,
while the supposedly ``improved'' formula works only at the few percent
level?

      That is the ``miracle'' and, but for a few remarks, I shall
leave the matter there for the reader to reflect upon.  There is much
more data than I have discussed here, including some for the O(2)-symmetric
case.  It all supports the same conclusion.  I can see no way that such
precise agreement could arise by accident, or from any kind of mistake.
All that need be said about the latter possibility is that the lattice
calculation is well-defined and should be eminently reproducible.

\vspace*{2mm}
\begin{center}
{\bf Acknowledgements}
\end{center}

\vspace*{-1.5mm}
I thank Maurizio Consoli for valuable discussions.

This work was supported in part by the U.S. Department of Energy under
Grant No. DE-FG05-92ER40717.

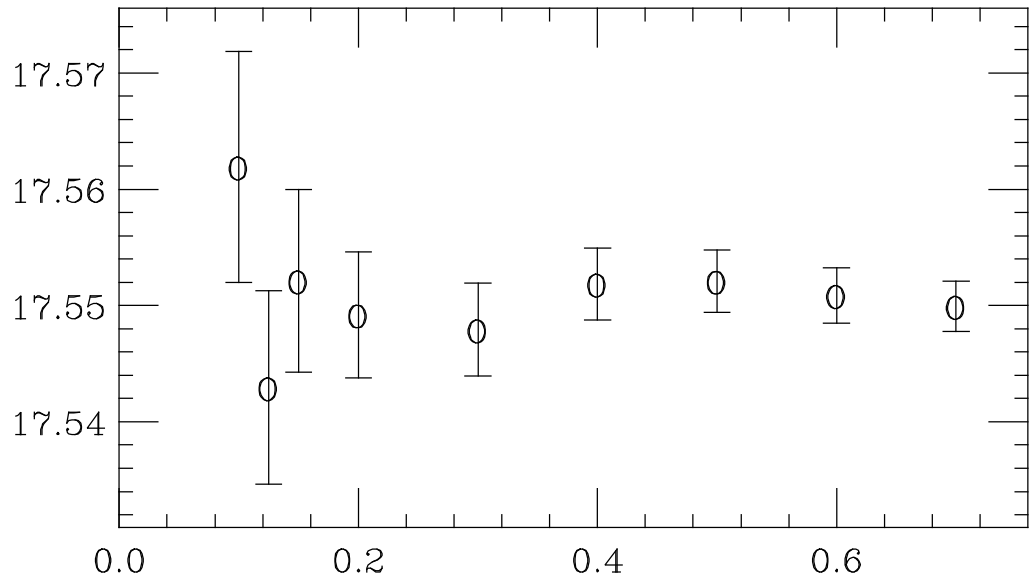
\end{document}